# The new concepts of measurement error's regularities and effect characteristics

Ye Xiaoming[1,2,*] Liu Haibo [3,4] Xiao Xuebin [5] Ling Mo[3]

[1] School of Geodesy and Geomatics, Wuhan University, Wuhan, Hubei, China, 430079.
[2] Key Laboratory of Precision Engineering & Industry Surveying, State Bureau of Surveying and Mapping, Wuhan, Hubei, China, 430079
[3] Institute of Seismology, China Earthquake Administration, Wuhan, Hubei China, 430071.
[4] Wuhan Institute of earthquake metrological verification and measurement engineering, Wuhan, Hubei China, 430071.
[5] Wuhan University Library. Wuhan, Hubei, China 430079.

Abstract: In several literatures, the authors give a new thinking of measurement theory system based on error non-classification philosophy, which completely overthrows the existing measurement concept system of precision, trueness and accuracy. In this paper, by focusing on the issues of error's regularities and effect characteristics, the authors will do a thematic interpretation, and prove that the error's regularities actually come from different cognitive perspectives, are also unable to be used for classifying errors, and that the error's effect characteristics actually depend on artificial condition rules of repeated measurement, and are still unable to be used for classifying errors. Thus, from the perspectives of error's regularities and effect characteristics, the existing error classification philosophy is still incorrect; and an uncertainty concept system, which must be interpreted by the error non-classification philosophy, naturally becomes the only way out of measurement theory.

## 1. Introduction

In several literatures [1] [2] [3], the authors give a new thinking of measurement theory system based on error non-classification philosophy. The main logic of this thinking is briefly introduced as follows:

The concept of error is defined as the difference between the measurement result and its true value. Because the measurement result is unique, and the true value is also unique, so the error of the measurement result is the only unknown and constant deviation.

For a final measurement result, the constant deviation consists of two parts: 1, the deviation $\Delta_A$ between the final measurement result and mathematical expectation, which is the so-called random error in existing theory; 2, the deviation $\Delta_B$ between mathematical expectation and true value, which is the so-called systematic error in existing theory. Because both deviations are unknown and persist constant deviations, and do not have any difference in characteristics, therefore, having no characteristic difference must not cause any classification difference!

The standard deviation of deviation $\Delta_A$ is given by the statistic and analysis of current measurement data. The deviation $\Delta_B$ is also produced by measurement; its formation principle is actually the same as the current measurement; its standard deviation can be obtained by tracing back to its upstream measurement. Thus, the standard deviation of total error of final measurement result is equal to the synthesis of the two standard deviations according to the probability laws. This total standard deviation is uncertainty, which is the evaluation of the probable interval of the error of final measurement result (this give a more clear meaning to the uncertainty concept).

This constant deviation theory is completely opposite to the random variation theory of existing measurement theory, that is, in the opinion of the authors, it is obviously illogical that existing measurement theory interpret deviation $\Delta_A$ as precision but interpret deviation $\Delta_B$ as trueness, and the error classification definition and all the concepts of precision, trueness and accuracy should be abolished.

For example: in 2005, the Chinese surveying and Mapping Bureau gave that the elevation result of Mount Everest is 8844.43 meters with standard deviation of $\pm 0.21$ meters. According to existing error classification theory, from the perspective of error's definition, the error of this result is a single constant deviation and should be classified as systematic error; however, from the perspective of standard deviation $\pm 0.21$m, it should be classified as random error. This is the logical trap of existing error classification theory. And the interpretation, according to error non-classification theory, is that this result's error (the difference between the result and the true value at implementing measurement) is an unknown constant, and that the standard deviation of $\pm 0.21$m is only the evaluation of the probable interval of the unknown constant error. That's it.

---
* xmye@sgg.whu.edu.cn

The difference between the new theory and the existing theory is shown in Fig1.

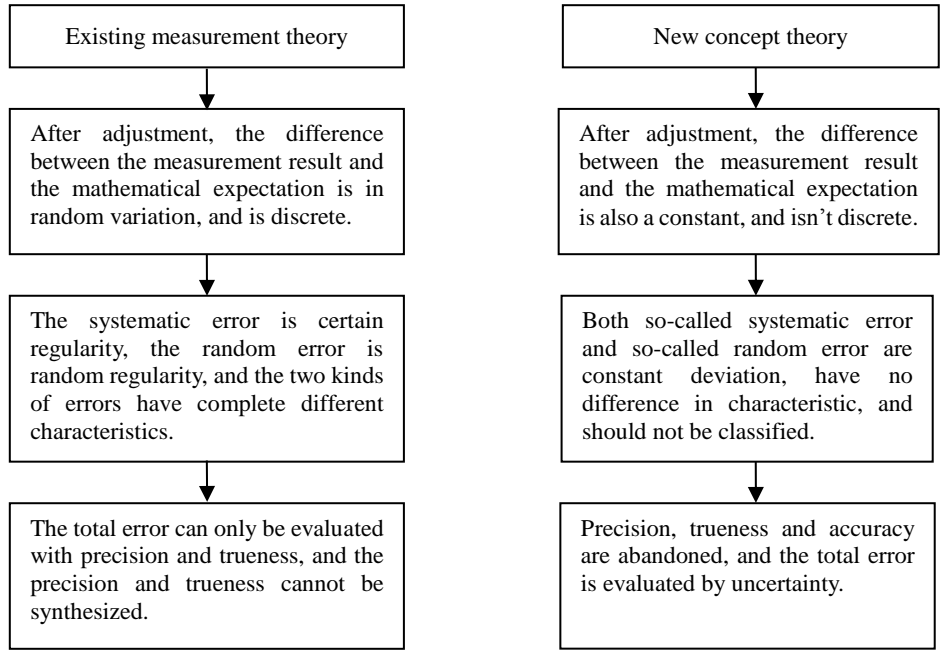

Fig1. The comparison of two theory's logic

Please note that the authors' emphasize on the concept of constant deviation is to focus on the measurement result instead of the original observation values before forming final measurement result. Of course, the authors recognize that there may be indeed a discrete error sample sequence before the measurement result is formed. However, these discrete error samples, which have certain numerical value, are the measured values of errors. They belong to the measurement result, and naturally cannot be mixed with the unknown error of final result to discuss the error classification. In addition, the dispersion and deviation of error sample sequences actually depend on the conditions of repeated measurements (Circuit noise is also a condition), naturally cannot be used to prove that the error can be classified.

The core difference between the two theories is, the existing theory considers that the error can be classified into systematic error and random error, while the new theory holds that the error has no systematic and random classification. Although document [1] has mentioned that the regularity and influence characteristics of errors cannot be used for classifying errors, the relationship between regularity and randomness, the formation mechanism of error's influence characteristics and related applications have not been interpreted in detail. Therefore, this paper will make a detailed interpretation on the regularity and influence characteristics of error.

## 2. Error's regularity

The concept of error is the difference between the measurement result and its true value. The error must be a constant deviation, that is to say, any single error is a constant.

The task of measurement theory is to study the methods of reducing and evaluating error. From the unknown and constant characteristics of single error, this task naturally faces difficulty. However, before the final measurement result is formed, our measurement is usually repeated, and there will be many error samples. When we observe a group of error samples, the errors can show some regularity, including certain regularity and random regularity. This provides paths for reducing and evaluating error: by certain regularity we can design some methods for compensating and correcting error; by random regularity we can design the statistic method for reducing error and obtain the evaluation method of error.

That is, the issue of error's regularity is actually aimed at a group of error samples before the final measurement result is obtained, instead of single error after the final measurement result is obtained.

However, it is important to note that the error's certain regularity and random regularity are actually from different perspectives. They are different error processing methods, and naturally cannot be used to achieve error classification. The same kind of error can be processed according to certain regularity, and also can be processed according to random regularity. There is still not error's classification issue according to certain regularity and random regularity. These are also the ideas from the new theory, which is totally different from the existing measurement theory.

For example: the measured frequency values of a quartz crystal at different temperatures are shown in Table 1.

According to Table 1, to observe the error values alongside the temperature values, we can get the certain regularity as shown in Fig2. However, the error value is observed alone, we can get random regularity as shown in Fig3.

That is to say, corresponding to the temperature values to observe the error values, we see the certain regularity; viewing the temperatures as arbitrary and only observing the error's distribution, we see the random regularity. Naturally, there are two ways to deal with it in practice.

Table 1. The measured frequency values of a quartz crystal at different temperatures

| Temperature °C | Frequency MHz | Error value $R_i = \Delta f_i / f_0 (\times 10^{-6})$ |
|---|---|---|
| -40° | 4.999900 | -30 |
| -30° | 4.999975 | -15 |
| -20° | 5.000040 | -2 |
| -10° | 5.000085 | 7 |
| 0° | 5.000115 | 13 |
| 10° | 5.000110 | 12 |
| 20° | 5.000070 | 4 |
| 30° | 5.000035 | -3 |
| 40° | 5.000010 | -8 |
| 50° | 4.999995 | -11 |
| 60° | 4.999995 | -11 |
| 70° | 5.000010 | -8 |
| 80° | 5.000045 | -1 |
| 90° | 5.000125 | 15 |
| 100° | 5.000235 | 37 |

1, Random model processing:

Error equation: $v_i = f_i - f_0$

According to the least square method, the final measurement result is:

$$f_0 = \frac{\sum_{i=1}^{n} f_i}{n} = 5.000050 MHZ$$

Its standard deviation:

$$\sigma_{\Delta f} = \sqrt{\frac{\sum_{i=1}^{n} v_i^2}{n-1}} = \pm 15.8 \times 10^{-6} f_0$$

That is, the frequency value of the quartz crystal is 5.000050MHZ, and its standard deviation (in the temperature between -40 and 100 degrees) is $\pm 15.8 \times 10^{-6}$. This expresses that the actual error of the frequency value exists in a probability interval with standard deviation $\pm 15.8 \times 10^{-6}$ at arbitrary temperature between -40 and 100 degrees.

2, Function model processing:

The function model of temperature-frequency error is $R = a + bT + cT^2 + dT^3$.

Error equation: $v_i = R_i - a - bT_i - cT_i^2 - dT_i^3$

According to the least square method, there is

$$\begin{pmatrix} n & \sum T_i & \sum T_i^2 & \sum T_i^3 \\ \sum T_i & \sum T_i^2 & \sum T_i^3 & \sum T_i^4 \\ \sum T_i^2 & \sum T_i^3 & \sum T_i^4 & \sum T_i^5 \\ \sum T_i^3 & \sum T_i^4 & \sum T_i^5 & \sum T_i^6 \end{pmatrix} \begin{pmatrix} a \\ b \\ c \\ d \end{pmatrix} = \begin{pmatrix} \sum R_i \\ \sum R_i T_i \\ \sum R_i T_i^2 \\ \sum R_i T_i^3 \end{pmatrix}$$

Substituting the values in Table 1 into above equation, there are:

$$\begin{pmatrix} 15 & 450 & 41500 & 292500 \\ 450 & 41500 & 2925000 & 256870000 \\ 41500 & 2925000 & 256870000 & 21952500000 \\ 292500 & 256870000 & 21952500000 & 1983295000000 \end{pmatrix} \begin{pmatrix} a \\ b \\ c \\ d \end{pmatrix} = \begin{pmatrix} -1 \\ 4610 \\ 304500 \\ 42713000 \end{pmatrix}$$

Solving the equations, get:
$a = 9.983251, b = -0.013518, c = -0.018601, d = 0.000214.$

Therefore, the frequency error's function model is fitted as:

$$R = 9.983251 - 0.013518T - 0.018601T^2 + 0.000214T^3$$

Fig4 is the comparison curve between the model and the actual error.
The standard deviation of residual error is

$$\sigma_{\Delta R} = \sqrt{\frac{\sum_{i=1}^{n} v_i^2}{n-4}} = \pm 2.3 \times 10^{-6}$$

Finally, the frequency of quartz crystal is given as follows:
$$f = f_0(1 + R \times 10^{-6})$$

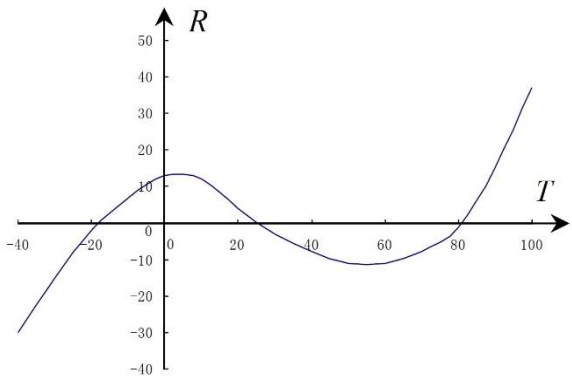

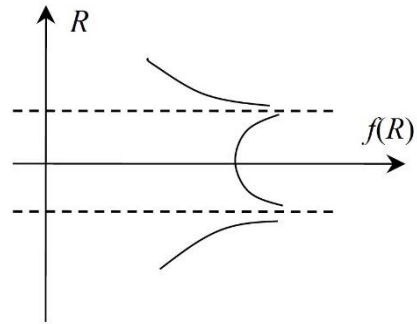

Fig2.The temperature-frequency error of quartz crystal

Fig3.The frequency error's distribution

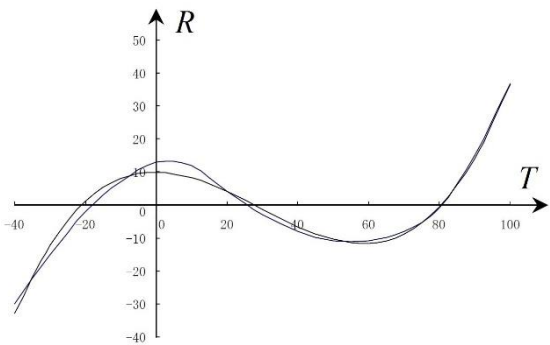

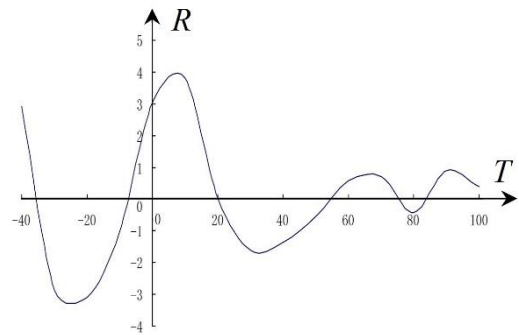

Fig4.Frequency error curve fitted by function model

Fig5.The residual error's curve

That is, temperature-frequency error can be corrected by the measurement value of temperature sensor, and a more accurate frequency value can be calculated. Residual error (as shown in Fig5) is still processed by statistical rules, and the standard deviation of the residual error is reduced to $\pm 2.3 \times 10^{-6}$. This error processing method has been widely used in the manufacture of photoelectric geodimeter [4][5].

Note that, although the effect of random model is not as good as the function model, it does not mean that the random model processing method is incorrect! In fact, at above function model processing, the final residual error (Fig5) is still processed by random model. However, it can be seen from the Fig5, the residual error is actually still a regular error rather than white noise, and has no essential difference with Fig2.

Another example, the cycle error of the phase type photoelectric geodimeter[4][5] shows periodic function regularity with distance. Its function model is: $y = A\sin\varphi$, but when the phase $\varphi$ is regarded as arbitrary, its probability density function is:

$$f(y) = \begin{cases} \dfrac{1}{\pi\sqrt{A^2 - y^2}} & (|y| \leq A) \\ 0 & (|y| > A) \end{cases}$$

Its standard deviation is:

$$\sigma_y = \frac{A}{\sqrt{2}}$$

It can be seen, when relating error $y$ with phase $\varphi$ to observe, the error shows sine regularity; when the phase $\varphi$ is viewed as arbitrary, the sine cycle error is also to follow a random distribution.

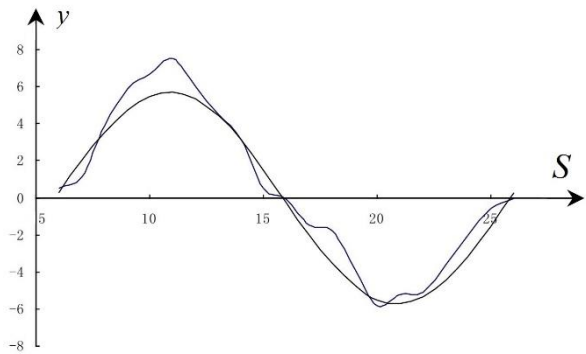 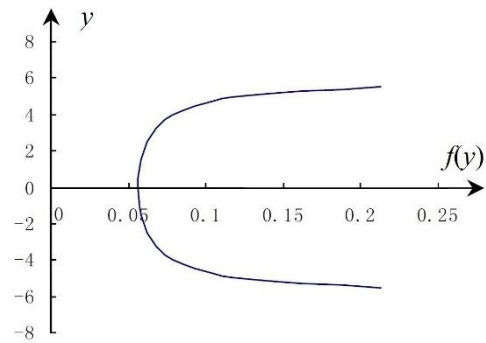

Fig6.The function model fitting of cycle error     Fig7.The cycle error's distribution

Table 2 is the testing data of an instrument. By the data of Table 2, the cycle error's function model is fitted as $y = 5.7\sin(\dfrac{S}{20} \times 360° + 254.41°)(mm)$ (see Fig6). Naturally, as shown in Fig7, its probability density function is:

$$f(y) = \begin{cases} \dfrac{1}{\pi\sqrt{5.7^2 - y^2}} & (|y| \leq 5.7mm) \\ 0 & (|y| > 5.7mm) \end{cases}$$

Table 2. The testing data of cycle error of a geodimeter

|   | Standard distance (m) | Measured distance (m) | Error value (mm) |
|---|---|---|---|
| 1 | 6.0237 | 6.0232 | 0.5 |
| 2 | 7.0239 | 7.0228 | 1.1 |
| 3 | 8.0243 | 8.0204 | 3.9 |
| 4 | 9.0246 | 9.0187 | 5.9 |
| 5 | 10.0250 | 10.0183 | 6.7 |
| 6 | 11.0253 | 11.0178 | 7.5 |
| 7 | 12.0256 | 12.0196 | 6.0 |
| 8 | 13.0258 | 13.0213 | 4.5 |
| 9 | 14.0263 | 14.0232 | 3.1 |
| 10 | 15.0266 | 15.0261 | 0.5 |
| 11 | 16.0269 | 16.0270 | -0.1 |
| 12 | 17.0269 | 17.0284 | -1.5 |
| 13 | 18.0269 | 18.0287 | -1.8 |
| 14 | 19.0268 | 19.0307 | -3.9 |
| 15 | 20.0267 | 20.0325 | -5.8 |
| 16 | 21.0268 | 21.0320 | -5.2 |
| 17 | 22.0269 | 22.0320 | -5.1 |
| 18 | 23.0269 | 23.0305 | -3.6 |
| 19 | 24.0270 | 24.0290 | -2.0 |
| 20 | 25.0271 | 25.0277 | -0.6 |
| 21 | 26.0272 | 26.0272 | 0 |

That is to say, to observe by relating the distance with the error value, we see the cycle regularity; viewing the distance as arbitrary and only observing error distribution, we see a random regularity. Naturally, in practice, there are also two methods to deal with it.

1．Random model processing:

Table 3 is the simulation data of using the cycle error $y = 5\sin(\dfrac{S}{20} \times 2\pi + \dfrac{\pi}{4})(mm)$ under the measured distance $S_{BC}$=8.0000m, and simulated the 15 groups distance difference data randomly and arbitrarily.

Error equation:  $v_i = S_{1i} - S_{2i} - S_0$

According to the least square method, the final measurement result is:

$$S_0 = \frac{\sum\limits_{i=1}^{n}(S_{1i} - S_{2i})}{n} = 8.0014(m)$$

Visible, the error is only 1.4*mm*, less than the amplitude of the cycle error.

2．Function model processing:

The function model of cycle error is:

$$y = A\sin(\frac{S}{\lambda} \times 2\pi + \phi) = a\sin\frac{S}{20} \times 2\pi + b\cos\frac{S}{20} \times 2\pi$$

The error equation is:

$$v_i = S_{1i} - y_{1i} - (S_{2i} - y_{2i}) - S_0$$
$$= S_{1i} - S_{2i} - a(\sin\frac{S_{1i}}{20} \times 2\pi - \sin\frac{S_{2i}}{20} \times 2\pi) - b(\cos\frac{S_{1i}}{20} \times 2\pi - \cos\frac{S_{2i}}{20} \times 2\pi) - S_0$$

Make

$$S_i = S_{1i} - S_{2i}, A_i = \sin\frac{S_{1i}}{20} \times 2\pi - \sin\frac{S_{2i}}{20} \times 2\pi, B_i = \cos\frac{S_{1i}}{20} \times 2\pi - \cos\frac{S_{2i}}{20} \times 2\pi.$$

The error equation become into:

$$v_i = S_i - A_i a - B_i b - S_0$$

According to the least square method, there is:

$$\begin{pmatrix} n & \sum A_i & \sum B_i \\ \sum A_i & \sum A_i^2 & \sum A_i B_i \\ \sum B_i & \sum A_i B_i & \sum B_i^2 \end{pmatrix} \begin{pmatrix} S_0 \\ a \\ b \end{pmatrix} = \begin{pmatrix} \sum S_i \\ \sum A_i S_i \\ \sum B_i S_i \end{pmatrix}$$

Replace the data in Table 3 into above equation:

$$\begin{pmatrix} 15 & 3.64249 & 2.39534 \\ 3.64249 & 27.83084 & -3.44106 \\ 2.39534 & -3.44106 & 26.44747 \end{pmatrix} \begin{pmatrix} S_0 \\ a \\ b \end{pmatrix} = \begin{pmatrix} 120.0214 \\ 29.22612 \\ 19.24404 \end{pmatrix}$$

There are:

$$S_0 = 8.00000, a = 0.00353, b = 0.00353$$

In this way, the amplitude of the cycle error is $A = \sqrt{a^2 + b^2} = 0.00499(m)$, and its phase is $\phi = \arctan\frac{a}{b} = \frac{\pi}{4}$.

Table 3. The simulation data of using cycle error

|   | $S_{AB}$ (m) | $S_{AC}$ (m) | $S_2 = S_{AB} + y_{AB}$ (m) | $S_1 = S_{AC} + y_{AC}$ (m) |
|---|---|---|---|---|
| 1 | 10 | 18 | 9.9965 | 18.0008 |
| 2 | 12 | 20 | 11.9951 | 20.0035 |
| 3 | 33 | 41 | 32.9951 | 41.0045 |
| 4 | 27 | 35 | 27.0008 | 34.9965 |
| 5 | 22 | 30 | 22.0049 | 29.9965 |
| 6 | 28 | 36 | 27.9992 | 35.9977 |
| 7 | 30 | 38 | 29.9965 | 38.0008 |
| 8 | 36 | 44 | 35.9977 | 44.0045 |
| 9 | 38 | 46 | 38.0008 | 46.0023 |
| 10 | 26 | 34 | 26.0023 | 33.9955 |
| 11 | 34 | 42 | 33.9955 | 42.0049 |
| 12 | 16 | 24 | 15.9977 | 24.0045 |
| 13 | 18 | 26 | 18.0008 | 26.0023 |
| 14 | 19 | 27 | 19.0023 | 27.0008 |
| 15 | 42 | 50 | 42.0049 | 49.9965 |
| $AC_0 - AB_0 =$ 8.0000 | | | | |

It can be seen, regular error follows random distribution, can be processed according to function model, and also can be processed according to random model. Although the effect of function model is actually better than the random model, the premise of using function model is that the function model is known. If the function model is unknown, we naturally think its regularity is an "unpredictable manner", and use the random model to process it. In the practice of measurement, it is a common fact that one or more regular errors are processed according to random model, and the so-called random regularity is more because their certain regularities are ignored or unknown.

For example: besides the cycle error, the multiplicative constant error of photoelectric geodimeter[4][5], which is the residual error after temperature corrected, is still the function of temperature(see Fig5), and called as systematic error by existing theory, but processed according to random model instead of temperature function model in traverse survey[6].

Another example: in the level[7][8], the i angle error, cross error, compensation error, focusing error, and so on, are the regular errors, and called the systematic error by existing theory, but processed according to random model in leveling network measurement[9].

Another example: steel ruler's thermal expansion error, watch's running error, gauge nominal value's error, also are the regular errors, but manufacturers usually only give these equipment's maximum permissible error (MPE) or total standard deviation indicator, which is actually also the random model processing.

Although these error handling methods in all the above cases already exist in measurement practice, it is clear that using random model to deal with these regular errors is obviously contrary to the concept logic of existing measurement theory. These also show that the conceptual interpretation based on the error classification in existing theory does not conform to the actual measurement.

When discussing random regularity, we have to discuss the electronic noise (white noise or 1/f noise), which is a random function of time and is an unavoidable error source in the field of electronic measurement, although most of measurements actually have no physical mechanism of electronic noise error. Because of having not mastered its regularity, the electronic noise error can only be processed with random model.

What is worthy of noting is, in the actual measurement, because the randomness of various process conditions drives one or more regular errors (also may include electronic noise) to randomly change, error sample sequence also shows random distribution (is similar to noise's random distribution). However, these discrete error samples aren't the random function of time, because the finish of measurement data collection has fixed all the measurement data, and all the data are unable to change with time. That is to say, after the data processing is completed, the error of final measurement result is unable to contain a component which randomly changes with time, and the contribution of noise to the final result's error is also a constant deviation.

That is, except the function model cannot be used, the electronic noise hasn't any essential particularity, and through white noise concept considering the error of final measurement result as random function of time is a misunderstanding.

In short, the error's regularity is an observation effect through observing a group of error samples instead of a single error; error's variation is certainly associated with the variations of measurement condition, and these measurement conditions are temperature, measurement range, instrument, time, location, leveling, sighting, electronic noise and so on; error's certain regularity and random regularity are observation results from different perspectives, they have no mutual exclusion, and taking regularities to achieve error classification is similarly impossible. It is obviously inappropriate that VIM [10][11] takes "predictable manner" and "unpredictable manner" to define the error classification.

## 3. Error's effect characteristics

The new theory considers error has no systematic and random classification. It refers to that the error has no difference whether it follows random distribution, but does not negative error can produce systematic or random effects. The error's effect characteristics and error's random distribution are two different things: following random distribution refers to that the error exists in a finite probable interval, but systematic or random effects refer to that the error sources contribute deviation or dispersion to subsequent repeated measurement. See Table 4.

Table 4. The comparison of concept of two theories

| Current theory | The new concepts theory |
| --- | --- |
| Error is classified as systematic error and random error. | Error cannot be classified according to systematic and random way. |
| The systematic error does not follow random distribution, and the random error follows random distribution. | Any error follows a random distribution. |
| Random distribution is random variation. | Random distribution is that the error is in a finite probable interval instead of random variation. |
| Systematic error contributes systematic effects (contribute deviation), and random error contributes random effects (contribute dispersion). | The error's systematic or random effects depend on the variation rules of measuring conditions in repeated measurements. |
| The systematic error is certain regularity, and the random error is random regularity. | The error's regularity depends on the perspectives of observation, and the error can show various regularities. |

It can be seen, the core of the existing systematic error concept is that it does not follow random distribution, but the new theory stresses that any error follows a random distribution and that the error's systematic effect is completely different from the concept of existing systematic error.

Just as important, the error's systematic or random effect depends on the variation rule of the measuring conditions in the repeated measurements, which is actually another angle of the error's regularity issue.

In the case of quartz crystal's frequency, the frequency error varies with the temperature, so the temperature is the related measurement condition. If the repeated measurements are in constant temperature, temperature - frequency error will remain unchanged, produce systematic effects, and not drive the observation sequence dispersion; if the repeated measurements are in different temperature, temperature - frequency error will change, produce random effects, and drive the observation sequence dispersion.

In the case of the photoelectric geodimeter[4][5], the cycle error is the periodic function of distance, so the distance is the related measurement condition. If repeated measurements are in the same distance condition, the cycle error will remain unchanged, produce systematic effects, and not drive the observation sequence dispersion; if repeated measurement is in different distance condition, the cycle error will change, produce the random effects, and drive the observation sequence dispersion. (Such as Table 3).

Moreover, besides systematic and random effect characteristics, error also has the characteristic of non-effect. For example, using the differential method to measure distance (as shown in Table 3), the additive constant error of photoelectric geodimeter[4][5] has no effect to the observations $S_i = S_{1i} - S_{2i}$. Also, all the errors, which have no intrinsic physical relation with the observation, are unable to affect the observation. For example: the error of instrument A cannot affect the observation of instrument B.

Hence, the error's systematic and random effects or observation sequence's deviation and dispersion depend on the changing rules of repeated measurement conditions. Temperature, measurement range, instrument, time, locations, leveling, even circuit noise, and so on, are measurement conditions. The same error can produce systematic effects in a repeated measurements, also can produce random effects in another repeated measurement, and even cannot produce effect. Naturally, using effect characteristics to classify error is still impossible. And it is a mistake that existing theory equate the error's effect characteristics with the error's classifications.

## 4. The new interpretation of uncertainty concept

Because the existing uncertainty [12] [13] [14] concept system accepts the error classification philosophy and uncertainty's definition clearly expresses the meaning of "dispersion", many people naturally understand it as being similar to precision. This kind of uncertainty, is neither fish nor fowl of course, naturally causes controversy [15].

Now, any regular error follows a random distribution and has its standard deviation. The theory of error classification is overthrown. Naturally, the concepts of precision, trueness and accuracy must be abolished, and an uncertainty concept system, which must be interpreted by the error non-classification philosophy, has become the only way out of measurement theory.

The total error of measurement result is a constant deviation and has no classification. Its numerical value is unknown, and uncertain. The unknown and uncertain degree of error's numerical value is the uncertainty, which is expressed by the evaluation of the probable interval of error. Because the total error comes from the synthesis of many error sources according to algebraic law, the total standard deviation is equal to the synthesis of standard deviations of all the source errors according to the covariance propagation law.

Further, because the numerical value of measurement result is certain and the numerical value of error is uncertain, the uncertainty also expresses the uncertain degree that the true value cannot be determined.

Uncertainty is the evaluation value of the probable interval of the error of measurement result, and expresses the degree that the true value cannot be determined or the probable degree that the measurements result is close to the true value. This is the new interpretation of the uncertainty concept.

In some cases, the variation of the true value in the future and the ambiguity of the true value definition also should be considered as the error problem, thus, a broad understanding of uncertainty is given.

A simple example for comparing: The indication error of photoelectric geodimeter[4][5] consists of the additive constant error $C$, the multiplicative constant error $R$, the periodic error $P$, and the dividing error δ, which are all residual errors after being processed by instrument. Now, the distance value given by the instrument is *S*. So, how does the error of the measurement result be evaluated?

1, According to the traditional error classification thinking, additive constant error *C* is a constant regularity, multiplicative constant error *R* is proportional regularity and is also a non-linear regularity of temperature (Fig5), and periodic error *P* is sine regularity, so they are all systematic errors. Only the regularity of the dividing error δ is not clear, which belongs to the random error and has its standard deviation $\sigma(\delta)$. Therefore, the precision of measurement result *S* is $\sigma(\delta)$ and express the dispersion of final measurement result, and the systematic errors *C*, *R* and *P* have no standard deviation, express the trueness of measurement result, which belongs to the category of qualitative evaluation.

However, because the uncertainty in the existing theory is also defined as dispersion, then what is the difference between uncertainty and precision? It is clear that this cannot be explained.

2, According to the concept logic of the new theory, the error of final measurement result is from the superposition of four deviations of *C*, *SR*, *P* and δ, and the error equation is

$$\Delta = C + S \cdot R + P + \delta$$

Because error has no classification and any error has its standard deviation, according to the covariance propagation law, the uncertainty of measurement result is

$$\sigma(\Delta) = \sqrt{\sigma^2(C) + S^2 \sigma^2(R) + \sigma^2(P) + \sigma^2(\delta)}$$

Among them, *C*, *R*, *P* and δ are all unknown errors, and their standard deviations are obtained by consulting the product specification of the instrument. The uncertainty $\sigma(\Delta)$ is the evaluation value of the probable interval that error $\Delta$ exists.

## 5. Conclusion

The single error of any final measurement result is constant regularity, and error classification cannot be achieved by the same constant regularity. Although a group of error samples can show some certain regularity and random regularity, certain regularity and random regularity are observation result from different perspectives and also cannot be used for classifying error. Because error's effect characteristics only depend on the variation rule of the measuring conditions in repeated measurements, and the same kind of error can show various kinds of effect characteristics, using effect characteristics to classify error is still impossible.

From all the perspectives, including single error's constant characteristic, the variation regularity of a group of error samples, and error's effect characteristics, classifying error cannot be realized. Naturally, the concept logic system of precision, trueness and accuracy of based on error classification theory shall completely collapse, and an uncertainty concept system, which must be interpreted by the error non-classification philosophy, has become the only way out of measurement theory.

Taking the same conditions in repeated measurements will make all the source errors to remain constant, hence cannot make error to be reduced. By appropriately changing the relevant measurement conditions in repeated measurements, any error can be made to contribute dispersion. Thus, error reduction can be realized by function model or random model processing, and the evaluation of probable interval of error also can be obtained. Uncertainty is the evaluation value of the probable interval of the error of measurement result, and expresses the probable degree that the final measurement result is close to the true value. This is the new interpretation of the uncertainty concept.